\documentclass[aip,twocolumn,letterpaper]{revtex4}

\usepackage{fullpage}

\usepackage{xcolor}
\definecolor{ultramarine}{rgb}{0.07, 0.04, 0.56}

\usepackage{graphics}
\usepackage{epsfig}

\usepackage[margin=1.0in]{geometry}

\begin{document}
\title{Electron Inertia and Magnetic Reconnection}
\author{Allen H Boozer}
\affiliation{Columbia University, New York, NY  10027 \linebreak ahb17@columbia.edu}

\begin{abstract}
When electron inertia is the only non-ideal effect in the evolution of a magnetic field $\vec{B}$, the field lines of $\vec{B}$ reconnect, but the lines of a related field $\vec{\mathcal{B}}$ do not.  $\vec{\mathcal{B}} \equiv \vec{B} + \vec{\nabla}\times \left( (c/\omega_{pe})^2\mu_0\vec{j} \right)$ with $\omega_{pe}$ the plasma frequency  and $\vec{j}$ the current density.   Although a full four-dimensional relativistic calculation of $\vec{\mathcal{B}}$ has been made, studies of  $\vec{\mathcal{B}}$ have been focused on systems that depend on only two spatial coordinates.  Three results are given:  (1)  A relatively simple demonstration in three dimensional space that the lines of $\vec{\mathcal{B}}$ do not reconnect  when electron inertia is the only non-ideal effect.  (2) The guiding center motion of charged particles is modified by a term that is proportional to $(c/\omega_{pe})^2$, which is smaller than the drifts proportional to the gyroradius unless the current density is extremely large.  (3) In three dimensional space, the evolution velocity of $\vec{\mathcal{B}}$ is characteristically chaotic, which means neighboring streamlines separate exponentially on a timescale $\tau_u$.  $\vec{\mathcal{B}}$ undergoes large scale reconnection on a timescale that is only an order of magnitude or two longer than $\tau_u$ unless all diffusive non-ideal effects, such as resistivity, are absolutely zero.

\end{abstract}

\date{\today} 
\maketitle


\section{Introduction}

This paper has three parts.  The first part, Section \ref{sec:non-recon}, shows that when electron inertia is the only non-ideal effect in the evolution of a magnetic field, then there is a divergence-free modified magnetic field
\begin{equation}
\vec{\mathcal{B}} \equiv \vec{B} + \vec{\nabla}\times \left( \frac{c^2}{\omega_{pe}^2}\mu_0\vec{j} \right) \label{Eq:modified-B}
\end{equation}
that cannot reconnect; $\omega_{pe}$ is the electron plasma frequency.    This result was given in 2015 by Lingam, Morrison, and Tassi \cite{Inertial-MHD:2015} using a two spatial-coordinate Hamiltonian-fluid method.  A full four-dimensional relativistic derivation was also published in 2015 by Asenjo and Comisso  \cite{Rel-e-recon:2015}.  Here a simple derivation of the three-coordinate result is given, which is independent of the Hamiltonian-fluid method.

Although much of the magnetic reconnection literature is based on two-coordinate models, two-dimensional theory is often misleading.   As discussed in Section \ref{Chaos-reconnection}, this is primarily because it is mathematically impossible for magnetic field lines to be chaotic throughout a finite volume when the magnetic field depends on only two coordinates.   Magnetic fields that depend on all three spatial coordinates are characteristically chaotic throughout volumes, which means the separation between infinitesimally separated lines has an exponential dependence on distance along the lines.  Anyone who has tried to design stellarator magnetic fields in three dimensions can testify about the difficulty of obtaining magnetic configurations in which the magnetic field lines lie on nested surfaces rather than chaotically filling a volume.

The second part of the paper, Section \ref{sec:implications}, is on the implications of the existence of a non-reconnecting magnetic field.   What is found is that the effects of the reconnection due to electron inertia on the particle guiding-center trajectories are proportional to $(c/\omega_{pe})^2$ and small compared to the effects linearly proportional to the gyroradius. Gyroradius effects dominate, even for the electrons, unless the current density is extremely large. 

The third part of the paper, Section \ref{Chaos-reconnection}, is on the applicability of the non-reconnection theorem of $\vec{\mathcal{B}}$ in three dimensional space.  In three dimensional space, the flow velocity of the lines of $\vec{\mathcal{B}}$ are naturally chaotic, which means neighboring streamlines separate exponentially on a timescale $\tau_u$.  Unless all non-ideal diffusive effects, such as resistivity, exactly vanish, the field  $\vec{\mathcal{B}}$ will undergo a large scale reconnection on a timescale within an order of magnitude or two of $\tau_u$.  The non-reconnecting nature of $\vec{\mathcal{B}}$ would seem to be of little practical importance in three dimensional systems.


\section{Proof that $\vec{\mathcal{B}}$ does not reconnect \label{sec:non-recon}}

Bruno Coppi \cite{Coppi:1964} in 1964 was the first to point out that the finite electron mass can produce magnetic reconnection even in the absence of collisions.     Any change in the current density in a plasma must be accompanied by an electric field to accelerate the current carriers, which gives a non-ideal evolution.   The importance of this effect has been discussed in many papers including \cite{Wesson:1990, Ottaviani:1993,  Biskamp, Shivamoggi, Andres:2014}.   The concept of inertial reconnection is a closely analogous to the London penetration depth of superconductors.  The form the reconnection takes can be viewed, Subsection \ref{Section:Voigt}, as a physical example of a Voigt renormalization \cite{Constantin:2023,Yi-Min:2025}, $\vec{B} \rightarrow \vec{B} + (c/\omega_{pe})^2\nabla^2\vec{B}$, which smoothes the magnetic field over the scale of plasma skin depth $c/\omega_{pe}$. 

In deriving the two-coordinate expression for the non-reconnecting field $\vec{\mathcal{B}}$, Equation (\ref{Eq:modified-B}),  Lingam et al \cite{Inertial-MHD:2015} used a Hamiltonian-fluid formulation and included a long list of historical references that begins with a 1982 paper by Morrison and Greene \cite{Morrison:1982} and includes the 1994 paper of T. J. Schep, F. Pegoraro, and B. N. Kuvshinov \cite{Schep:1994}.    Other papers studying  reconnection due to electron inertia in two-coordinate Hamiltonian-fluids include the 1998 paper E. Cafaro, D. Grasso, F. Pegoraro, F. Porcelli, and A. Saluzzi \cite{Ham-recon:1998} and the 2001 paper by Grasso, Califano, Pegoraro, and Porcelli \cite{Grasso:2001}.  

The method that is used in this section to demonstrate that  $\vec{\mathcal{B}}$ does not reconnect in three-coordinate systems is not based on the theory of Hamiltonian fluids.  Instead, a simple direct proof is based on the general mathematical condition for reconnection for any field that evolves in accordance with an equation of the form of Faraday's Law.

This proof will be given in three subsections.  Subsection \ref{sec: mathcal B} obtains the expression for the divergence-free field $\vec{\mathcal{B}}$ of Equation (\ref{Eq:modified-B}) and shows that the time evolution of this field is given by an equation of the Faraday's Law form, Equation (\ref{Mod-Faraday}).  The effective or modified electric field $\vec{\mathcal{E}}$ is given in Equation (\ref{First math-E}).

Subsection \ref{sec:simplification} derives a simplified form for $\vec{\mathcal{E}}$, Equation (\ref{Mathcal-E}). 

Subsection \ref{sec:recon} derives the condition that is required for reconnection to occur in any divergence-free field that obeys an equation of the form of Faraday's Law.  This subsection also shows the component of the modified electric field $\vec{\mathcal{E}}$ along  $\vec{\mathcal{B}}$ can be written as the gradient of a single-valued potential, which implies the field lines of $\vec{\mathcal{B}}$ flow with a velocity $\vec{u}_\bot$ and cannot change topology.


\subsection{Modified magnetic field \label{sec: mathcal B}}

In the frame of reference in which the ions are stationary, the current density $\vec{j}$ and the electron force balance equation are
\begin{eqnarray}
&& \vec{j} = -en\vec{v}_e;\\
&& m_e n\frac{d \vec{v}_e}{dt} = -en(\vec{E}+\vec{v}_e\times\vec{B}) \nonumber \\  &&\hspace{0.7in} - \vec{\nabla}\cdot\tensor{P}_e -m_en\nu_{ei} \vec{v}_e,  \hspace{0.2in} \mbox{so  } \label{electron}\\
&& \vec{E} = \frac{\vec{j}\times\vec{B}}{en}  - \frac{\vec{\nabla}\cdot\tensor{P}_e}{en} + \frac{ m_e \nu_{ei}}{e^2n} \vec{j} + \frac{d}{dt}\left( \frac{m_e}{ne^2} \vec{j} \right).   \hspace{0.2in}
\end{eqnarray}
The resistive term $\eta\vec{j}$ with $\eta \equiv m_e \nu_{ei}/(ne^2)$ and the electron pressure/viscosity term $\vec{\nabla}\cdot\tensor{P}_e/en$ cause reconnection and are set to zero in order to study the effect of electron inertia.  

Expanding $d\vec{v}_e/dt =\partial \vec{v}_e/\partial t + \vec{v}_e\cdot \vec{\nabla}\vec{v}_e$, reconnection will be studied when the electric field is
\begin{eqnarray}
&& \vec{E}=   \frac{\vec{j}\times\vec{B}}{en}+ \frac{\partial}{\partial t}\left( \frac{m_e}{n e^2} \vec{j} \right) -\frac{\vec{j}\cdot\vec{\nabla}}{en_e} \left( \frac{m_e}{n e^2} \vec{j} \right) , \hspace{0.2in}\\ 
&&\mbox{where} \hspace{0.2in} \frac{m_e}{n e^2} \vec{j}= \frac{c^2}{\omega_{pe}^2} \mu_0\vec{j};
\end{eqnarray} $c/\omega_{pe}\approx 0.5~$cm when the electron density is $10^{20}~$m$^{-3}$.

The evolution of the magnetic field is given by Faraday's Law, $\partial \vec{B}/\partial t = - \vec{\nabla}\times \vec{E}$ so
\begin{eqnarray}
 &&\frac{\partial}{\partial t}\left(\vec{B} +  \vec{\nabla}\times\left( \frac{m_e}{n e^2} \vec{j}\right)\right) = -\vec{\nabla}\times\vec{\mathcal{E}} \\
&&\vec{\mathcal{E}} \equiv \frac{\vec{j}\times\vec{B}}{en}-\frac{\vec{j}\cdot\vec{\nabla}}{en} \left( \frac{m_e}{n e^2} \vec{j} \right). \label{First math-E}
\end{eqnarray}

Although electron inertia breaks the lines of $\vec{B}$, it does not break the lines of the divergence-free field
\begin{eqnarray}
&& \vec{\mathcal{B}} \equiv \vec{B} + \vec{\nabla}\times \left( \frac{m_e}{n e^2} \vec{j} \right) \label{def-math-B} \hspace{0.2in} \mbox{with   } \\
&& \frac{\partial \vec{\mathcal{B}}}{\partial t}= - \vec{\nabla} \times \vec{\mathcal{E}} \label{Mod-Faraday}.
\end{eqnarray}


\subsection{Simplified form for $\vec{\mathcal{E}}$ \label{sec:simplification}} 

Equation (\ref{First math-E}) for $\vec{\mathcal{E}}$ can be rewritten in  a simplified form   
\begin{eqnarray}
\vec{\mathcal{E}} &=& \frac{\vec{j}\times\vec{\mathcal{B}}}{en} -  \frac{e}{2m_e} \vec{\nabla}\left( \frac{m_e\vec{j}}{n e^2}  \right)^2.  \label{Mathcal-E}
\end{eqnarray}

The derivation of the simplified form starts with replacing $\vec{B}$ by $\vec{\mathcal{B}}$ in the Hall term.
\begin{equation}
\frac{\vec{j}\times\vec{B}}{en} = \frac{\vec{j}\times\vec{\mathcal{B}}}{en} -  \frac{\vec{j}}{en} \times (\vec{\mathcal{B}}-\vec{B}).
\end{equation}
The term
\begin{eqnarray}
&& \frac{m_e\vec{j}\times(\vec{\mathcal{B}}-\vec{B}) }{ne^2} \nonumber \\ &&
 \hspace{0.2in}= \left(\frac{m_e\vec{j}}{ne^2}\right)\times \left(\vec{\nabla}\times  \frac{m_e\vec{j}}{n e^2}  \right) \\
 && \hspace{0.2in}= \frac{1}{2} \vec{\nabla}\left( \frac{m_e\vec{j}}{n e^2}  \right)^2   - \left( \frac{m_e\vec{j}}{n e^2}  \right)\cdot\vec{\nabla}\left( \frac{m_e \vec{j}}{n e^2} \right). \hspace{0.2in}
\end{eqnarray}
Consequently the Hall term has the form
\begin{eqnarray}
\frac{\vec{j}\times\vec{B}}{en} &=& \frac{\vec{j}\times\vec{\mathcal{B}}}{en}-\frac{e}{2m_e} \vec{\nabla}\left( \frac{m_e\vec{j}}{n e^2}  \right)^2  \nonumber\\ && + \left( \frac{\vec{j}}{n e}  \right)\cdot\vec{\nabla}\left( \frac{m_e \vec{j}}{n e^2} \right), 
\end{eqnarray}
which implies Equation (\ref{First math-E}) for  $\vec{\mathcal{E}}$ can be rewritten as Equation (\ref{Mathcal-E}).


 \subsection{Reconnection condition \label{sec:recon}}
 
 The proof that Equation (\ref{Mathcal-E}) for $\vec{\mathcal{E}}$ implies that the lines of $\vec{\mathcal{B}}$ cannot reconnect starts with a purely mathematical equation, the representation of the field $\vec{\mathcal{E}}$ in terms of $\vec{\mathcal{B}}$. 
 
 Anywhere that $\vec{\mathcal{B}}\neq0$, mathematics implies any vector $\vec{\mathcal{E}}$ in three-space can be represented as 
\begin{equation}
\vec{\mathcal{E}} + \vec{u}_\bot \times \vec{\mathcal{B}} = - \vec{\nabla}\Phi +\mathcal{E}_r\vec{\nabla}\ell,  \label{E eq}
\end{equation}
where $\ell$ is the the distance along a field line of $\vec{\mathcal{B}}$ with $\vec{\mathcal{B}}\cdot\vec{\nabla}\ell=\mathcal{B}$, and $\bot$ means perpendicular to $\vec{\mathcal{B}}$.  

The potential $\Phi$ can always be chosen so $\mathcal{E}_r$ is constant along $\vec{\mathcal{B}}$. When $\mathcal{E}_r$ can be chosen to be zero the evolution of $\vec{\mathcal{B}}$ is ideal, and the field lines are carried with the velocity $\vec{u}_\bot$, the field line velocity, as shown by Newcomb \cite{Newcomb}  and below. 

Replacing $\vec{\mathcal{E}}$ in Equation (\ref{Mod-Faraday}) by $\vec{\mathcal{E}}$ from  Equation (\ref{E eq}) demonstrates that the evolution of $\vec{\mathcal{B}}$ is given by an advection-diffusion equation \cite{Aref:1984,Tang-Boozer:96}.  When the advective term is chaotic, advection-diffusion equations have an exponential sensitivity to the diffusion \cite{Tang-Boozer:96}.  An exception is when $\vec{\mathcal{B}}$ is limited to two spatial dimensions.  Then a chaotic $\vec{u}_\bot$ gives an exponential increase in $\mathcal{B}^2$ \cite{Boozer:2025}.

The validity of Equation (\ref{E eq}) is proven by showing that all three components of $\vec{\mathcal{E}}$ can be represented.  The parallel component is
\begin{eqnarray}
\frac{ \vec{\mathcal{B}} }{\mathcal{B}} \cdot \vec{\mathcal{E}} & =& -  \frac{ \vec{\mathcal{B}} }{\mathcal{B}}\cdot\vec{\nabla}\Phi + \mathcal{E}_r\\
\frac{ \vec{\mathcal{B}} }{\mathcal{B}} \cdot\vec{\nabla}\Phi &=& \left( \frac{\partial\Phi}{\partial\ell}\right)_{\alpha\beta},  \hspace{0.2in} \mbox{so   }\\
\mathcal{E}_{||} &=& - \left( \frac{\partial\Phi}{\partial\ell}\right)_{\alpha\beta} + \mathcal{E}_r
\end{eqnarray}  
using Clebsch coordinates $(\alpha,\beta,\ell)$. Since $\vec{\mathcal{B}}$ is divergence free, it can be written in the Clebsch form $\vec{\mathcal{B}}=\vec{\nabla}\alpha\times \vec{\nabla}\beta$.

Equation (\ref{Mathcal-E}) for $\vec{\mathcal{E}}$ then implies 
\begin{eqnarray}
\frac{\partial \Phi}{\partial \ell} &=& \frac{e}{2m_e} \frac{\partial }{\partial\ell}\left( \frac{m_e\vec{j}}{n e^2}  \right)^2,
\end{eqnarray}
which gives a single-valued expression for $\Phi$, so $\mathcal{E}_r$ can be taken to be zero.  The perpendicular components of  Equation (\ref{E eq}) are fit by $\vec{u}_\bot$. 

What is essentially Newcomb's proof \cite{Newcomb} that $\mathcal{E}_r=0$ ensures that the lines of  $\vec{\mathcal{B}}$ are carried by the flow, $\vec{u}_\bot$, follows from Faraday's Law.  Using the Clebsch representation,    $\partial \vec{\mathcal{B}}/\partial t =  \vec{\nabla}\times ( (\partial\alpha/\partial t)\vec{\nabla}\beta - (\partial\beta/\partial t)\vec{\nabla}\alpha)$.  Since $\vec{u}\times\vec{\mathcal{B}}=(\vec{u}_\bot\cdot\vec{\nabla}\beta)\vec{\nabla}\alpha  - (\vec{u}_\bot\cdot\vec{\nabla}\alpha)\vec{\nabla}\beta$, when $\mathcal{E}_r=0$, the evolution is consistent with $d\alpha/dt \equiv \partial\alpha/\partial t + \vec{u}_\bot\cdot \vec{\nabla}\alpha =0$ and $d\beta/dt \equiv \partial\beta/\partial t + \vec{u}_\bot\cdot \vec{\nabla}\beta =0$.  A field line is defined by giving $\alpha$ and $\beta$, and these labels of a field line are carried by the flow $\vec{u}_\bot$.  

\begin{figure}
\centerline{ \includegraphics[width=3.2 in]{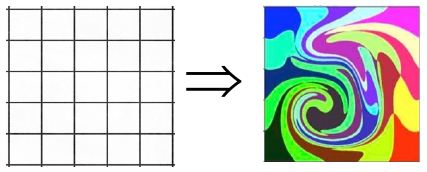}}
\caption{ A magnetic field $\vec{B}(\vec{x},t)$ can be thought of as consisting of tubes of magnetic flux by placing a gridded surface across the field.  Each tube is defined by the magnetic field lines that pass through the perimeters of the grid cells.  When the field is chaotic, the perimeter of each cell becomes exponentially longer when the grid is replotted after each line on the perimeters is followed for a distance $\ell$.   But, each cell contains exactly the same field lines and has precisely the same neighboring cells.  When the magnetic field is evolving ideally with a chaotic velocity $\vec{u}_\bot$, a similar distortion of the grid occurs when the grid is replotted using the location of each line on the perimeters after a time $t$.  The figure shows the distortion of a $5\times5$ array.   This is Figure 1 of Boozer, Phys. Plasmas \textbf{32}, 052106 (2025).  The distorted grid is part of Figure 5 of Y.-M. Huang and A. Bhattacharjee, Phys. Plasmas 29, 122902 (2022), which was based on a chaotic evolution defined by A. H. Boozer and T. Elder, Phys. Plasmas \textbf{28}, 062303 (2021).  Boozer and Elder illustrated distortions of ideally evolving flux tubes up to a factor $\sim 10^7$.  } 
\label{fig:tubes}
\end{figure}

 When the magnetic field lines are carried by a flow $\vec{u}_\bot$, field lines that lie in separate tubes of magnetic flux remain separated forever.  But as illustrated in Figure \ref{fig:tubes}, the surfaces bounding tubes of magnetic flux can become extremely contorted, which makes the preservation of distinct flux tubes hypersensitive to diffusive effects such resistivity.   See the discussion of Equation (\ref{B-ad-diff}).



\section{Implications \label{sec:implications}}


\subsection{Reconnection due to electron inertia}

This section shows that the effects on the the guiding center motion due to electron-inertia caused reconnection are second order in $c/\omega_{pe}$ while the effects on the guiding center motion due to gyroradius are first order.  Unless the current density $\vec{j}$ is extremely large, gyroradius effects are much larger than the effects due to reconnection caused by the electron inertia, even for the electrons.

The purely mathematical Equation (\ref{E eq}) applies to the magnetic field $\vec{B}$ as well as to $\vec{\mathcal{B}}$ once obvious notational changes have been made. The term $\mathcal{E}_r\vec{\nabla}\ell$ can be replaced in a torus by $(V_{\ell}/2\pi)\vec{\nabla}\varphi$ with $\varphi$ the toroidal angle.  The loop voltage $V_{\ell}$ gives the topological changes in the field.

 The electric field is
\begin{equation}
\vec{E} =  \frac{\partial}{\partial t}\left( \frac{c^2}{\omega_{pe}^2} \mu_0\vec{j} \right) + \vec{\mathcal{E}}
\end{equation}
To lowest order in the short distance $c/\omega_{pe}$, the parallel component of the electric field is given by
\begin{eqnarray} 
E_{||} &=& \frac{\vec{B}}{B} \cdot \frac{\partial}{\partial t}\left( \frac{c^2}{\omega_{pe}^2} \mu_0 \vec{j} \right).
\end{eqnarray} 
The loop voltage on an irrational magnetic surface or along any of the closed magnetic field lines of a rational surface is
\begin{eqnarray}
V_{||} &=& \frac{1}{M_t} \int_0^{2\pi M_t} E_{||} \frac{B}{\vec{B}\cdot\vec{\nabla}\varphi} d\varphi \hspace{0.2in}\mbox{as   } M_t\rightarrow\infty \hspace{0.2in} \\
&=&  \frac{1}{M_t} \int_0^{2\pi M_t} \frac{\vec{B}}{\vec{B}\cdot\vec{\nabla}\varphi} \cdot \frac{\partial}{\partial t}\left( \frac{c^2}{\omega_{pe}^2} \mu_0 \vec{j} \right) d\varphi,
\end{eqnarray}
which is not generally zero in an evolving magnetic field.


\subsection{Particle trajectories}

The importance of magnetic field lines comes in large part from their effect on the trajectories of charged particles.  When particles have a gyroradius $\rho \equiv v_\bot/\omega_c$ that is small compared to the size of the system, $a$, charged particles gyrate about $\vec{B}$ in near-circles of radius $\rho$.  The velocity of the particle perpendicular to the magnetic field is $\vec{v}_\bot$, and the cyclotron frequency is $\omega_c\equiv qB/m$ with $q$ the charge and $m$ the mass of the particle.  

When $|\rho/a|<<1$ and the magnetic and electric fields are slowly varying compared to $\omega_c$, the quantity $\mu\equiv mv_\bot^2/2B$ is an accurately conserved adiabatic invariant.  Averaging over the near-circular motion, the trajectory of a charged particle is accurately approximated by $d\vec{x}/dt=\vec{v}_g$, where $\vec{v}_g$ is known as the guiding center velocity.  In time-independent magnetic and electric fields
\begin{eqnarray}
\vec{v}_g = \frac{v_{||}}{B} \vec{\nabla}\times (\vec{A} + \rho_{||}\vec{B}) \label{v_g}  \hspace{0.2in} \mbox{with  }\\
\rho_{||}(H,\mu,\vec{x}) \equiv \frac{v_{||}}{\omega_c} \hspace{0.2in} \mbox{and  } \\
H = \frac{1}{2} v_{||}^2 + \mu B + q\Phi.
\end{eqnarray}
The full equation was given by Boozer \cite{Drift:Boozer}  in 1980; the perpendicular part by Morozov and L. S. Solov'ev \cite{Morozov:1966} in 1966.

The vector potentials of $\vec{B}=\vec{\nabla}\times\vec{A}$ and $\vec{\mathcal{B}}=\vec{\nabla}\times \vec{\mathcal{A}}$ are related by
\begin{eqnarray}
\vec{\mathcal{A}} = \vec{A} +  \frac{m_e}{n e^2} \vec{j}.
\end{eqnarray}
In a two-dimensional problem with no dependence on the $z$ coordinate, the $z$ component of canonical momentum of the electron motion, $P_z = m_e v_{ez} + (-e)\vec{A}$ is conserved.  M. Lingam et al \cite{Inertial-MHD:2015} proved that for this case $\vec{\mathcal{A}}= P_z/(-e)$ is conserved, and there can be no reconnection of the $\vec{\mathcal{B}}$ field.   When the magnetic field has $z$ dependence, no components of the canonical momentum need to be conserved, but the field $\vec{\mathcal{B}}$ still cannot reconnect.

Using the vector potential $\vec{\mathcal{A}}$ in Equation (\ref{v_g}), the guiding center velocity is 
\begin{eqnarray}
\vec{v}_g = \frac{v_{||}}{B} \vec{\nabla}\times \left(\vec{\mathcal{A}} + \rho_{||}\vec{B} - \frac{m_e}{n e^2} \vec{j}  \right) \label{math v_g}.
\end{eqnarray}
This equation can be used to obtain the guiding center trajectories relative to the modified field, which has no reconnection due to electron inertia rather than relative to $\vec{B}$, which does reconnect due to electron inertia.

Comparing Equation (\ref{Mathcal-E}) and Equation (\ref{E eq}), the velocity $\vec{u}_\bot$ of the lines of $\vec{\mathcal{B}}$ is
\begin{equation}
\vec{u}_\bot = \frac{\vec{\mathcal{B}}}{\mathcal{B}^2} \times \left( \frac{e}{2m_e} \vec{\nabla}\left( \frac{m_e\vec{j}}{n e^2}  \right)^2 -\frac{\vec{j}\times\vec{\mathcal{B}}}{en} \right).
\end{equation}
The calculation was done in the frame of reference in which the ions are stationary.  The field lines could be taken to be stationary if the ions moved with a velocity $-\vec{u}_\bot$.

  When the current density becomes extremely high by being concentrated in thin current sheets, the flow must be essentially parallel to the magnetic field lines to avoid $\vec{j}\times\vec{B}$ forces that cannot be balanced.  As will be seen in Section \ref{Section:Voigt}, when $\left| (c/\omega_{pe})^2 \nabla^2\vec{B}\right| << \left|\vec{B}\right|$, the difference between $\vec{\mathcal{B}}$ and $\vec{B}$ is small.  To lowest order in  $c/\omega_{pe}$ and $\rho_{||}$, the guiding center velocity can be approximated as
\begin{eqnarray}
\vec{v}_g = \frac{v_{||}}{B} \vec{\nabla}\times \left(\vec{\mathcal{A}} + \left(\rho_{||}- \frac{c^2}{\omega_{pe}^2} \frac{\mu_0 j_{||}}{B}\right)\vec{\mathcal{B}}  \right) \label{math v_g}.
\end{eqnarray}
The ion gyroradius in a tokamak power plant will be comparable to $c/\omega_{pe}\approx 0.5~$centimeters and the electron gyroradius is about sixty times smaller.  Since $\mu_0 j_{||}/B \approx 1/qR$, where $q$ is the safety factor and $qR\approx 20~$m, the dimensionless ratio $(c/\omega_{pe})(\mu_0j_{||}/B)\approx 2.5\times10^{-4}$.  The $\mu_0j_{||}/B$ term would seem to have a negligible effect on even the electron trajectories unless very thin current layers are formed.


\subsection{Voigt-regularized magnetic field \label{Section:Voigt}}

The relationship between $\vec{\mathcal{B}}$ and the Voigt-regularized magnetic field follows simply when the number density $n$ is constant in space and time.  The generalized Ampere's law is
\begin{eqnarray}
\mu_0\vec{j}=\vec{\nabla}\times\vec{B} - \mu_0\epsilon_0\frac{\partial \vec{E}}{\partial t},
\end{eqnarray}
where $\mu_0\epsilon_0=1/c^2.$
When the spatial and temporal variations in the electron density are negligible, Equation (\ref{def-math-B}) is
\begin{eqnarray} 
\vec{\mathcal{B}}-\vec{B} &=&  \frac{c^2}{\omega_{pe}^2} \vec{\nabla}\times \mu_0\vec{j} \\
&=&  \frac{c^2}{\omega_{pe}^2} \left(-\nabla^2\vec{B} - \frac{\partial}{\partial t} \vec{\nabla}\times\vec{E}\right) \\
&=&- \frac{c^2}{\omega_{pe}^2} \Big(\nabla^2\vec{B} - \frac{1}{c^2} \frac{\partial^2 \vec{B}}{\partial t^2}\Big)
\end{eqnarray}
using Faraday's Law.  The relativistically invariant Laplacian is $\nabla^2\vec{B} - (1/c^2)\partial^2 \vec{B}/\partial t^2.$


\section{Chaos and the applicability of the non-reconnection of $\vec{\mathcal{B}}$ \label{Chaos-reconnection} }

Magnetic field line chaos fundamentally changes the theory of magnetic reconnection, Figure \ref{fig:tubes}.   The effect of chaos on election-inertia-produced reconnection in specific magnetic geometries was investigated in four papers by Borgogno and collaborators \cite{Borgogno:2005, Borgogno:2008, Borgogno:2011a, Borgogno:2011b}  from 2005 to 2011.     In their 2005 paper,  Borgorno et al carried out a numerical study of collisonless but unstable plasmas that initially had nested magnetic surfaces.  This means spatially bounded constant-$\psi(\vec{x},t)$ surfaces with $\vec{B}\cdot\vec{\nabla}\psi=0$.  If the magnetic evolution were perfectly ideal, the constant-$\psi(\vec{x},t)$ surfaces could be deformed but not destroyed.  Nevertheless, the weak non-ideal effects present in a collisionless plasma can cause these surfaces to break and create spatial volumes in which the magnetic field lines become chaotic.  In volumes of space that contain both constant-$\psi(\vec{x},t)$ surfaces and chaotic regions, Borgorno et al \cite{Borgogno:2005} measured the reconnected flux by the sum of the magnetic fluxes in all of the chaotic regions that are bounded by the remaining constant-$\psi(\vec{x},t)$ surfaces.  This reconnected flux was found to increase exponentially in time at approximately the instability growth rate once the instability amplitude was sufficient to produce chaotic regions.  Borgogno et al stated in their paper \cite{Borgogno:2011a}: ``The development of magnetic field line reconnection in three-dimensional configurations is an intrinsically chaotic process.''  Although true, this is recognized by only a remarkably small fraction of those studying magnetic reconnection.

The importance of magnetic field line chaos to understanding the behavior of magnetic fields and plasmas was discussed by Rosenbluth et al \cite{Rosenbluth:1966} in 1966.  The requirement for three spatial coordinates for magnetic field line chaos became clear in 1983 after Boozer showed \cite{Boozer:1983} that magnetic field lines are given by a magnetic field line Hamiltonian of the $H(p,q,t)$ form in canonical coordinates.  Unlike the standard textbook examples of Hamiltonians of the $H(p,q,t)$ form, the the three coordinates $(p,q,t)$ do not have an obvious physical interpretation in terms of momentum, position, and time.  The three Cartesian spatial coordinates $(x,y,z)$ are functions of  $(p,q,t)$.  Clock time is not a canonical variable in the field-line Hamiltonian; it is a parameter.  When a magnetic field depends on only two spatial coordinates, its Hamiltonian is of the $H(p,q)$ form, which implies $H$ is an invariant along each trajectory and an exponential separation of neighboring trajectories throughout a non-zero region of space is not mathematically possible.  A separatrix is possible with a X-line (confusingly commonly called a X-point) from which field lines exponentially separate, but a given large exponential separation can only occur in a region that is exponentially small by the same factor.  

A general mathematical theory of chaos and magnetic reconnection involves concepts such a cantori and turnstiles, which are not broadly known in the plasma physics community but were recently reviewed in \emph{Magnetic Field Line Chaos, Cantori, and Turnstiles in Toroidal Plasmas}, \cite{Boozer:chaos}.     Cantori, although not by that name, were discussed in two papers \cite{Borgogno:2011a, Borgogno:2011b} published by Borgorno et al in 2011.  They noted that ``certain surfaces may act as barriers to this magnetic field line penetration," the penetration of magnetic field lines through surfaces near which they linger. 

Simple derivations of effects associated with magnetic reconnection were given in the 2025 article, \emph{Magnetic reconnection and dynamos in the presence of plasma turbulence}, \cite{Boozer:2025}.  In particular, it is shown that a chaotic flow of the magnetic field lines causes an exponential increase in the magnetic field energy in two-dimensional systems---the magnetic field lines themselves cannot become chaotic.  In three-coordinate systems, a chaotic field-line flow gives chaotic magnetic field lines.

The fundamental point is that the evolution of any field that obeys an equation of the form of Faraday's Law can be written using Equation (\ref{E eq}) as an advection-diffusion equation.  A less general demonstration is to insert the simple Ohms' Law coupled with Ampere's Law, $\vec{E}+ \vec{v}\times\vec{B} =(\eta/\mu_0)\vec{\nabla}\times\vec{B}$, into Faraday's Law to obtain
\begin{eqnarray} \frac{\partial \vec{B}}{\partial t} - \vec{\nabla}\times(\vec{v}\times\vec{B} ) &=&- \vec{\nabla}\times \left( \frac{\eta}{\mu_0} \vec{\nabla}\times \vec{B} \right) \\
&=& \frac{\eta}{\mu_0} \nabla^2\vec{B}, \label{B-ad-diff}
\end{eqnarray}
when the diffusion coefficient $\eta/\mu_0$ is a spatial constant.  A general theory of magnetic reconnection follows from the properties of the advection-diffusion equation.

H. Aref revolutionized the theory of mixing by his 1984 paper \emph{Stirring by chaotic advection} \cite{Aref:1984}.  He pointed out that when the advective velocity is chaotic, neighboring streamlines have separations that depend exponentially on time, and that complete mixing---even at the molecular level---occurs on a time scale that only depends logarithmically on the diffusion coefficient as that coefficient approaches zero.  His argument was intuitive---essentially showing a figure related to Figure \ref{fig:tubes}.  However, Tang and Boozer \cite{Tang-Boozer:96} gave a rigorous proof of exponentially enhanced mixing of scalar quantities, such as density, using Lagrangian coordinates in 1996.  They also gave \cite{Tang-Boozer:2000}  the related solution to Equation (\ref{B-ad-diff}) for the magnetic evolution in 2000.  A 2002 and a 2005 publication \cite{Boozer:2002,Boozer:2005} are interesting historical references on chaos and magnetic reconnection

In the absence of any non-ideality in the magnetic evolution other than electron inertia, the field $\vec{\mathcal{B}}$ of Equation (\ref{Eq:modified-B}) cannot reconnect even when $\vec{\mathcal{B}}$ depends non-trivially on all three spatial coordinates.   However, when an evolving modified field $\vec{\mathcal{B}}$ depends non-trivially on all three spatial coordinates, its flow speed $\vec{u}_\bot$ will naturally be chaotic, which means neighboring streamlines of $\vec{u}_\bot$ have separations that depend exponentially on time.  Let $\tau_u$ be the characteristic timescale of that exponentiation.  Then, if there are any diffusive reconnection effects, which have a timescale $\tau_d$, large scale reconnection of $\vec{\mathcal{B}}$ will occur on a characteristic timescale $\tau_u \ln(\tau_d/\tau_u)$.  For realistic near-ideal systems this natural logarithm is of order ten.  It is difficult to conceive of a physical situation that would make this logarithm a hundred.  Even the natural logarithm of the ratio of the age of the universe divided by the time it takes light to cross the width of a proton is $\approx 95$.  The current density $\vec{j}$ need only be enhanced by a factor of order the same natural logarithm before large scale reconnection occurs.


\section*{Acknowledgements}

The author would like to thank Peter Constantin, the John von Neumann Professor of Mathematics at Princeton University, who was sufficiently intrigued by the result that he sent me an alternative derivation.  The author would also like to thank Amitava Bhattacharjee, who suggested reference \cite{{Ham-recon:1998}}, and  Andrew Brown, who pointed out reference \cite{Inertial-MHD:2015}.  Most importantly, he would like to thank Luca Comisso, who pointed  out the relativistic derivation of $\vec{\mathcal{B}}$.

This work received no external support.

 \vspace{0.01in}

\section*{Author Declarations}

The author has no conflicts to disclose. \vspace{0.01in}


\section*{Data availability statement}

Data sharing is not applicable to this article as no new data were created or analyzed in this study.


\end{document}